# Introduction to Focus Issue: Patterns of Network Synchronization

Daniel M. Abrams[1], Louis M. Pecora[2], and Adilson E. Motter[3]

[1]Department of Engineering Sciences and Applied Mathematics and Northwestern Institute on Complex Systems (NICO), Northwestern University, Evanston, Illinois 60208, USA

[2]U.S. Naval Research Laboratory, Washington, District of Columbia 20375, USA

[3]Department of Physics and Astronomy and Northwestern Institute on Complex Systems (NICO), Northwestern University, Evanston, Illinois 60208, USA

The study of synchronization of coupled systems is currently undergoing a major surge fueled by recent discoveries of new forms of collective dynamics and the development of techniques to characterize a myriad of new patterns of network synchronization. This includes chimera states, phenomena determined by symmetry, remote synchronization, and asymmetry-induced synchronization. This Focus Issue presents a selection of contributions at the forefront of these developments, to which this introduction is intended to offer an up-to-date foundation.

**Synchronization is an inherently collective dynamics phenomenon most suitably investigated within the framework of networks of coupled dynamical entities. Observed in a range of systems and conditions, synchronization manifests itself in a multitude of different scenarios, ranging from complete synchronization of all variables of all entities, to partial synchronization of some variables of some entities, to long lived but transient synchrony, to various forms of generalized synchronization. As in many other network phenomena, the nature of the collective dynamics depends on the properties of the entities, properties of the interactions (structure and dynamics), system size, coupling with external factors, and initial conditions. Characterizing the new, and often surprising, forms of synchronization that might emerge as a result of the interplay between these various factors is a major topic of current research. Below we give an introduction that attempts to show the various trajectories in dynamical systems research that led to the current interests in patterns of synchronization dynamics in networks. We do not intend to present a comprehensive review, but rather to give the reader a sense of the research paths that converged to this issue.**

In 1967 Arthur Winfree published a study on the modeling of biological rhythms and the behavior of populations of coupled oscillators [1], which inadvertently inspired volumes of research not only in biology but also in the physical, mathematical, and engineering sciences. In particular, it motivated Yoshiki Kuramoto to introduce in 1975 his celebrated solvable model [2], which established a firm connection between the onset of synchronization in coupled oscillators and phase transitions in statistical physics. By the time Winfree wrote his paper, the phenomenon of synchronization had been know for at least 300 years, with the original discovery often attributed to Christiaan Huygens' 1665 fortuitous observation of synchronization of pendulum clocks [3]—incidentally the same year Robert Hooke reported his discovery of the living cell [4]. Its pertinence to biological systems had also long been known. George Mines, for example, whose premature death at age 28 apparently resulted from experimentation with self-induce fibrillation, had developed by as early as 1914 experimental techniques to investigate cardiac arrhythmias resulting from loss of synchrony in the heart induced by small electric stimuli [5]. More recently, in 1959, Boris Belousov published on his observations of long-lived patterns of oscillatory dynamics in a nonlinear chemical reaction system [6]. While previous research had been largely observational and qualitative, the theme of Winfree and Kuramoto's work was theoretical and quantitative.

The legacy of those early studies combined with the ever increasing availability of computer resources and, more recently, concepts from network science, has led to significant new insights over the past few decades. It is now well established that systems of coupled entities can exhibit various forms of synchronization phenomena. This is the case for many types of coupled systems and forms of interactions, whether the entities are conscious, animate, or inanimate; periodic, aperiodic, or even chaotic; oscillatory, pulse-like, or excitatory; whether the system is of finite size or in the thermodynamic limit; whether entities are coupled locally, globally, or through a more general network; forming a discrete or continuous medium; described by iterated maps or differential equations, as deterministic or stochastic systems; coupled in isolation or driven by external chemical, thermal, optical, electrical, or mechanical stimuli; whether the interactions are pairwise or through higher-order (hypernetwork) structures; whether the coupling is weak or strong, static or time dependent; whether the entities and interactions are all identical or heterogeneous. Various forms of synchronization phenomenon have been shown, with their own particularities, in each of these scenarios.

What is then left to be discovered? From a basic science standpoint, a major outstanding question is to determine how different patterns of synchronization (including yet-to-be discovered ones) relate to the properties of the entities, the properties of the interactions (both internal and external), and the properties of the initial conditions. The significance of this question has become even more evident by a series of discoveries made over the past 15 years. For example, the discovery of chimera states—spatiotemporal patterns of coexisting synchronized and incoherent populations in networks of identically coupled identical oscillators [7, 8]—shows that multi-stability and symmetry breaking can lead to counterintuitive behavior not anticipated in previous studies of global synchronization. The discovery of relay and remote synchronization [9, 10], where mutually synchronized oscillators are connected through oscillators in asynchronous states, exposes yet another class of subtle collective behavior. The latter as well as numerous other patterns of cluster synchronization can now be systematically understood using a recently developed framework that relates stable cluster synchronization with symmetries in the underlying network of interactions [11]. More recently, synchronization in a network of identically coupled Stuart-Landau oscillators was used to demonstrate the converse of symmetry breaking: a scenario in which the system cannot synchronize when the oscillators are identical but synchronizes stably when the oscillators' parameters are tuned to be suitably different from each other [12]. As these examples abundantly illustrate, the relationship between the various factors that influence synchronization and synchronization dynamics is anything but obvious.

Because the concept of synchronization lies at the heart of many network dynamical patterns, we offer below a little more insight into the history of two main types of synchronization (phase and identical), which lead up to the papers in this Focus Issue. We then briefly summarize the content of the issue.

**PHASE SYNCHRONIZATION**

Phase synchronization (PS) is most often discussed in the context of phase oscillators, which in the absence of coupling satisfy the extremely simple linear ordinary differential equation $d\varphi_i/dt = \omega_i$: the $i$th oscillator increases its phase $\varphi_i$ at a constant rate $\omega_i$. Given an ensemble of such oscillators, coupling may induce synchronization such that all phases become equal at some point or points in time. That is, a state we refer to as phase-synchronized: $\varphi_i = \varphi_j$ for all $i$, $j$. Most often, PS (sometimes referred to as phase lock) refers to a *persistent* state in which all oscillators converge to the same (usually time-varying) phase. The related concept of frequency lock (confusingly, sometimes also called phase lock) refers to a state where coupling induces oscillators to converge to the same instantaneous frequency $d\varphi_i/dt$, but not necessarily the same phase. These concepts can also be applied, sometimes with slight modification, to more complicated oscillators. Important examples include oscillators with nonconstant $\omega_i$ (nonautonamous or nonlinear oscillators), oscillators with nonconstant amplitudes (higher-dimensional oscillators), and oscillators with higher-order derivatives in their differential equations (e.g., oscillators with inertia).

The foundational work in the study of PS was done by Yoshiki Kuramoto beginning in the 1970s, when he developed what has come to be known as the Kuramoto model [2] while studying nonequilibrium statistical mechanics. His reduction to phase oscillations and ingenious solution of that simplified model created a new paradigm for understanding collective behavior in nature. Much of the history of the research on PS is tied up with the study of the Kuramoto model; Steven Strogatz's 2000 review paper [13] does an excellent job of summarizing progress through that point and highlighting open questions. Dörfler and Bullo [14] updated progress in 2014.

In 1986, Kuramoto and his student Hidetsugu Sakaguchi generalized the Kuramoto model to allow for arbitrary coupling networks and the presence of a phase-lag term. This formulation is now often referred to as the Kuramoto-Sakaguchi model [15]. Exploration of that model first focused primarily on the case of discrete lattice networks (e.g., [16–18]), often motivated by oscillations in arrays of Josephson junctions. As interest in complex networks grew, more attempts were made to understand the impact of a variety of structures on synchronization properties [e.g., 19–24]. Several papers in this Focus Issue continue to examine the interplay between network structure and synchronization dynamics.

In 2002, Kuramoto and Dorjsuren Battogtokh published a paper showing the existence and apparent stability of a state with broken spatial symmetry in a ring of nonlocally coupled oscillators [7]. Two years later, motivated by the surprising combination of synchrony and incoherence within a fully symmetric system, Daniel Abrams and Steven Strogatz dubbed this a chimera state [8] and attempted to explain its origin. Much work has since been done to understand properties of these so-called chimera states; a review article published last year [25] summarizes some of it, but research is ongoing, and many papers in the current issue address chimera states.

The Kuramoto and Kuramoto-Sakaguchi models have motivated a variety of generalizations that continue to be of interest, including, for example, delay-coupling, unidirectional coupling, non-pairwise coupling, and heterogeneity in phase lags. These systems, along with comparisons to experimental data, comprise several more of the papers in our Focus Issue.

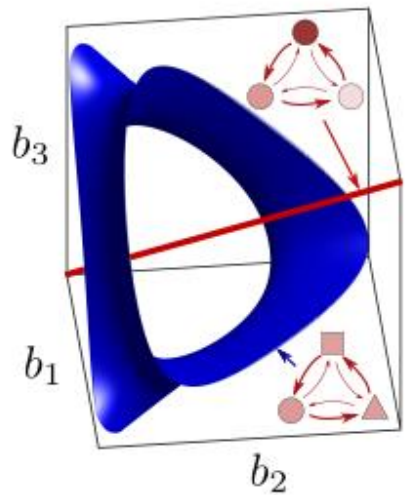

**Figure 1**. Symmetric state requiring system asymmetry in three-oscillator networks. Inside the blue surface is the region of stable synchronization in the three-dimensional parameter space, in which all nodes exhibit identical dynamics. Since all networks with identical node parameters $b_1$, $b_2$, and $b_3$ (illustrated by identical node shapes) lie on the red line, which does not intersect the blue region, the nodes can achieve identical dynamics only if the nodes themselves are non-identical. In this example, the nodes are amplitude-phase oscillators—a class that includes Stuart-Landau oscillators as special cases. (Adapted from [12].)

A final twist in the study of PS has come with the recent discovery by Takashi Nishikawa and one of us (AEM) of networks of amplitude-phase oscillators that can only synchronize identically when the oscillators themselves are not identical [12]. This is interesting because it has been generally assumed that individual entities are more likely to exhibit the same or similar behavior if they are equal to each other (e.g., think of lasers pulsing at the same frequency, birds singing the same notes, agents reaching consensus). But as the example in Fig. 1 shows, this assumption is generally false in oscillator networks. In its strongest form, this phenomenon can be shown for *zero-lag* synchronization and oscillators identically coupled in the network, meaning that the state of the system can be symmetric only when the system itself is not symmetric. For this reason, the phenomenon was termed *asymmetry-induced symmetry* and can be seen as the converse of the well-studied phenomenon of symmetry breaking.

**IDENTICAL SYNCHRONIZATION**

Generally when we speak of identical synchronization (IS) we mean two or more coupled identical dynamical systems whose dynamics are exactly the same; that is, they follow the same state space trajectory (for IS among non-identical systems, see Ref. [12]). This can mean they are phase or time lagged, but the same trajectory is followed for all coupled systems or, perhaps, certain subsets of a larger network.

A surprising form of IS comes from the identical synchronization of chaotic systems. This is initially counterintuitive since positive Lyapunov exponents tend to push two nearly synchronized oscillators apart in state space exponentially fast in time. The first papers we know of to show that chaotic IS is possible are those of Hirokazu Fujisaka and Tomoji Yamada [26–28]. They actually considered simple networks of identical systems coupled diffusively through all their dynamical variables. These were followed closely, but sporadically, by several other papers out of the Soviet Union. Arkady Pikovsky coupled two chaotic oscillators and showed in 1984 that two attractors could synchronize their motion [29]. A little later a now-famous paper appeared by Valentin Afraimovich, Nikolai Verichev, and Mikhail Rabinovich [30], which showed that two Duffing-like systems could exactly synchronize when linked by diffusive coupling. Note that early Soviet papers often referred to chaotic motion as stochastic motion. This can lead to some confusion and unfortunate exclusion of the works by others familiar with only the "chaotic" adjective. Finally, work by Alexander Volkovskii and Nikolai Rul'kov [31] closed out the decade of the 1980's with another take on IS.

In 1990 one of us (LMP) along with Thomas Carroll [32] found a new way to synchronize two oscillators in a one-way driving scheme. The goal was to eventually think of the system as a communication setup with the drive system being the transmitter and the driven system (also called the response) as the receiver. It was this potential application and the possibility that some type of privacy or even encryption could be added using the chaos that led soon after to a growing research effort in investigating IS and a growing body of literature. The topic of synchronizing a few coupled (usually chaotic) systems still shows up in different guises in the literature at this time. However, up to here there are no real networks or dynamics patterns in the system as a whole except for the synchronized motion of the few coupled oscillators. This changed in two stages, one with the introduction of networks to the synchronization scene and the other with the discovery of cluster synchronization.

Early work on global synchronization in networks of diffusively coupled systems of many oscillators is presented in the paper by Heagy et al. [33]. Although this was still global synchronization some of the techniques that emerged would later morph into general analysis tools. More sophisticated approaches appear in work by Gang Hu et al. [34], where it was shown how to deal with more general networks and parameter variations in their effects on global synchronization. Another important paper at this time was by Prashant Gade [35], who looked at much more complicated networks of randomly coupled maps. Some general notions of global synchronization, such as looking at the eigenvalues of the coupling matrix as depending on the number of connections in the network, showed that the two are related in some classes of networks [36]. Finally, in 1998 Pecora and Carroll [37] developed an analysis technique for closely linking the eigenvalue spectrum of the coupling matrix to the linear stability of the identical oscillator systems that works well for any type of dynamical motion (time-independent, periodic, quasiperiodic, and chaotic). This allowed for the separation of the network structure from the dynamics of the node oscillators, which simplified the problem since the latter only needed to be calculated once as what is now called the master stability function. Then a change in network only meant a change in eigenvalue spectrum and the stability of the system could be determined directly with no more dynamical calculations. These papers and the many that came after as well as the blossoming field of network science [38, 39] inspired much synchronization in networks research from 1998 up to the present.

The concept of clusters of synchronized nodes appearing in a network followed a few years later. This is the early work that led to many of the papers in this Focus Issue. Typical early works are analyzed and/or demonstrated in Refs. [10, 40–50]. These and other papers in this time period were very interesting, but mostly for special networks which were constructed to exhibit cluster synchronization intentionally. Earlier work in the famous book by Martin Golubitsky and colleagues [51] essentially showed that networks of identical oscillators could have cluster synchronization (although it was not called that) if the network had symmetries. Symmetries imply that certain subsets of nodes in the network can be permute among themselves, which implies immediately that the subset of nodes can have IS dynamics, although its dynamics will, in general, differ from nodes in other subsets. The subsets are then cluster patterns. The main problem is that one has to know the symmetries beforehand. This is also true for a recent publication on remote synchronization [10], which can be seen as symmetry-induced cluster synchronization, although the latter did develop some parts of a scheme to calculate the stability of the clusters. More recently a very general scheme for discovering cluster synchronization was developed [11]. The approach uses computer efficient discrete algebra software [52] to find the symmetries of a network, which are called automorphisms of the adjacency matrix. This allows one to study networks for which symmetries are not obvious or are so numerous that they cannot be counted by a human. Moreover, the machinery of group representations is also available in the software and this allows for the block diagonalizaton of the variational equations, greatly simplifying the calculation of the stability of the clusters.

Finally, another approach to finding cluster synchronization in networks was developed in [53] and earlier papers (see also the book by Golubitsky and Stewart [54]). In this approach it is assumed that nodes have the same intrinsic dynamics and whether two nodes will synchronize depends only on the inputs to them. If they have the same number of inputs from the same synchronized clusters (not necessarily the same nodes), they will synchronize. This approach is applied in Kevin Judd's work [49] and in O'Clery et al.'s [50]. The input approach is more general than the symmetry approach in finding synchronization clusters since there exist in some systems synchronization clusters that are not symmetry clusters. And there do exist efficient algorithms that will find the minimal number of input/synchronization clusters [55], although the problem of finding sub-synchronous clusters from there is a very hard, unsolved problem. However, in the calculation of stability of the clusters, the input method does not fully block diagonalize the variational equations like the symmetry method since powerful computational machinery of irreducible representations of groups has, so far, no counterpart in the graph theory associated with the input approach. At this time it is best to say that the symmetry and input methods complement each other.

Many of the above-mentioned authors have contributed to this Focus Issue. The reader now has a good chance to see where all these studies have led to today and get a good sense of what the new, unsolved problems are.

**THIS FOCUS ISSUE**

Here we briefly comment on the papers in this issue. The contributions cover a wide range of topics—a tribute to the diversity of research in the field—as reflected, for example, in the following papers:
- Ottino-Löffler and Strogatz [56] report on localized patterns in a lattice of Kuramoto oscillators, which are characterized by the instantaneous frequencies of the oscillators and manifest themselves as rotating spirals.
- Wang and Chen [57] introduce a metric-topological model to characterize global synchronization of moving agents on a plane.
- Lu et al. [68] consider a synchronization-based model to explain the asymmetry of human response to jet lag caused by eastward versus westward travel.
- Bick et al. [59] study the emergence of chaos in coupled oscillator networks that may involve simultaneous interaction of more than two oscillators.
- DeVille and Ermentrout [60] study the Kuramoto model on special sparse networks, and find that a variety of stable non-synchronized states can exist.
- Fujiwara et al. [61] study conditions for synchronization in time-varying networks of coupled oscillators.

Many papers in the issue can be partially grouped into (nonexclusive) themes. In particular, a number of contributions cover the topic of chimera states:
- Hart et al. [62] experimentally observe chimera states in a minimal network of four globally coupled identical opto-electronic oscillators with time-delayed feedback loops.

- Kemeth et al. [63] propose a new scheme for defining and classifying a wide range of chimera states.
- Martens et al. [64] generalize a two-cluster system that has previously been used to study chimera states by allowing for distinct phase lags within and across clusters.
- Belykh et al. [65] study a model similar to one often studied in the context of chimera states, but generalize it by allowing for distinct numbers of oscillators in each node and by giving the oscillators inertia.
- Nkomo et al. [66] report experimental chimera states in a system of coupled Belousov–Zhabotinsky chemical oscillators, and compare their experimental results with numerical simulations of a mathematical model.
- Ulonska, et al. [67] study Van der Pol oscillators coupled via networks of increasingly hierarchical nature, and find that large network clustering coefficients can promote the existence of chimera states of different types.

Several other contributions explicitly consider implications of symmetries on synchronization patterns:
- Golubitsky and Stewart [68] survey, within the formalism of coupled cell systems, symmetry and other mechanisms for patterns of synchrony and phase locking.
- Nishikawa and Motter [69] study the properties of networks optimized for the global synchronizability of coupled oscillators, and find that such networks exhibit symmetrical structures and can also support cluster synchronization.
- Schaub et al. [70] apply the graph-theoretical concept of external equitable partitions to find conditions for cluster synchronization to occur in general systems of Pecora-Carroll coupled oscillators.
- Sorrentino and Pecora [71] examine a network of coupled oscillators with small parameter mismatches both in the oscillator properties and network couplings, and develop low-dimensional methods for predicting synchronization error.

There are also several contributions (in addition to some above) that address problems of optimization and/or control in synchronization dynamics:
- Skardal et al. [72] study optimization of network structures for synchronization of coupled phase oscillators in directed networks with a given distribution of natural frequencies.
- Nagao et al. [73] experimentally study the use of delay-coupled networks of electrochemical reactions to show that changes in coupling can induce changes in behavior from oscillation death to anti-phase synchronization.
- Skardal and Arenas [74] analyze two methods of control for Stuart-Landau coupled oscillators, with the goal of directing the system to a desired state with a minimal intervention.
- Deng et al. [75] study the optimization of coupling strengths to suppress amplitude death in a chain of coupled Stuart-Landau oscillators.
- Tandon et al. [76] propose a control strategy in which chaotic nonidentical oscillators can be synchronized by temporarily uncoupling them when they enter certain regions of the phase space.

Finally, a number of other contributions focus on the impact of time delays, phase lags, and noise:
- Laing studies [77] the impact of delays on the existence of stable travelling waves in non-locally coupled networks of theta neurons and phase oscillators.
- Kobayashi and Kori [78] show that, in networks of noisy phase oscillators, synchronizability is reduced as degree heterogeneity is increased and that both noise and strong coupling tend to induce phase slip desynchronization..
- Focusing on the impact of phase lags on the synchronization transitions in the Kuramoto model, Omel'chenko and Wolfrum [79] show that there are scenarios for which synchrony may decrease for increasing coupling strength.
- D'Huys et al. [80] examine networks of Boolean switches with time-delayed coupling both experimentally and analytically, finding surprisingly long transients that scale exponentially with the time delay.
- Emenheiser et al. [81] investigate the deterministic and stochastic dynamics of nearest-neighbor-coupled micromechanical resonators, and characterize the switching that can occur between distinct attracting states.

**ACKNOWLEDGEMENTS**

We thank all the authors of this Focus Issue for their invaluable contributions. We apologize to colleagues who feel their papers should also be mentioned in this brief introduction—a comprehensive review would take many more pages.